\documentclass[journal=ancac3,manuscript=article]{achemso}
\usepackage[dvipsnames]{xcolor}
\usepackage{amsmath}
\usepackage{amssymb}
\usepackage{textcomp}
\usepackage{longtable}
\usepackage{graphics}
\usepackage{gensymb} 
\usepackage{graphicx}
\usepackage{epsfig}
\usepackage{epstopdf}
\usepackage{soul}
\setkeys{acs}{maxauthors=60}
\setkeys{acs}{etalmode=truncate}
\setkeys{acs}{articletitle = true}
\usepackage[utf8]{inputenc}
\usepackage{xcolor}
\usepackage{float}
\usepackage{hyperref}

\newcommand{\UAMT}{Departamento de F\'{i}sica Te\'{o}rica de la Materia
                  Condensada, Universidad Aut\'{o}noma de Madrid,
                  28049 Madrid, Spain}
\newcommand{\IFIMAC}{Condensed Matter Physics Center (IFIMAC),
                     Universidad Aut\'{o}noma de Madrid, 28049 Madrid, Spain}
\newcommand{\ICMM}{Instituto de Ciencia de Materiales de Madrid, 
                   Consejo Superior de Investigaciones Cient\'{i}ficas,
                   28049, Madrid, Spain}
\newcommand{\UT}{Department of Thermal and Fluid Engineering, 
                   University of Twente,
                   7500 AE, Enschede, The Netherlands}
\newcommand{\UCB}{Department of Chemical and Biological Engineering, 
                  University of Colorado Boulder, Boulder, CO 80309}

\author{O. Mateos-Lopez}
\affiliation{\ICMM}
\alsoaffiliation{\UAMT}

\author{J. G. Vilhena}
\affiliation{\ICMM}

\author{I. Armstrong}
\affiliation{\UCB}

\author{H. Heinz}
\affiliation{\UCB}

\author{M. Muñoz Rojo} 
\affiliation{\ICMM}
\alsoaffiliation{\UT}
\email{m.m.rojo@csic.es} 

\author{J. C. Cuevas}
\affiliation{\UAMT}
\alsoaffiliation{\IFIMAC}
\email{juancarlos.cuevas@uam.es} 

\title{\texorpdfstring{Stereochemical Vacuum Gap Explains Out-of-Plane Thermal Insulation in MXenes}{Stereochemical Vacuum Gap Explains Out-of-Plane Thermal Insulation in MXenes}}
\begin{document}
\newpage

\cleardoublepage

\begin{abstract}

Two-dimensional MXenes are promising materials for thermal management and  spectral camouflage, combining low 
out-of-plane thermal conductivity with low infrared emissivity and mechanical robustness. Yet the 
near-order-of-magnitude spread in experimental out-of-plane thermal conductivity measurements 
(0.14--0.8~W\,m\textsuperscript{$-1$}K\textsuperscript{$-1$}) and the systematic overestimation by simulations 
point to a fundamental gap in our understanding of heat transport in these materials.
Here, we argue these differences originate in the overlooked role of heterogeneous surface terminations. 
Using Non-Equilibrium Molecular Dynamics simulations of Ti\textsubscript{3}C\textsubscript{2}T\textsubscript{\textit{x}},
we show that this discrepancy arises from a stereochemically induced vacuum gap between adjacent layers, formed 
when surface terminations of different sizes coexist. Even minor deviations from homogeneous terminations drastically
suppress out-of-plane thermal conductivity, bringing simulated values into quantitative agreement with experiment.
We also show that thermal conductivity scales strongly with the atomic density, and that introducing bulky surface 
terminations, including residual water, reduces the thermal conductivity to ${\sim} \, 0.3$~W\,m\textsuperscript{$-1$}K\textsuperscript{$-1$}, 
an order of magnitude below homogeneous termination values and below the minimum thermal conductivity limit predicted for disordered solids. Thus, we propose a chemistry-driven route to engineer thermal 
transport in MXenes.
\end{abstract}

\cleardoublepage

Two-dimensional (2D) transition metal carbides (MXenes) \cite{Naguib_2011, Naguib_2012, Naguib_2014} are attracting attention for applications in infrared thermal management \cite{YangLi_2021, Gogotsi_2023, Gogotsi_2025}, thermal camouflage \cite{Shen_2023, LeiLi_2021} and energy storage \cite{Anasori_2017, Pang_2019}, due to a combination of low infrared emissivity \cite{YangLi_2021, Han_2023} and high mechanical resilience \cite{Lipatov_2020} with the anisotropic thermal transport properties of 2D materials \cite{Gogotsi_2025}.
In addition, MXenes' thermal properties are also defined by their rich surface chemistry. 
Surface terminations attached to each layer, such as O, F or OH groups \cite{Naguib_2013, Halim_2014, Halim_2016, Gogotsi_2025}, are determined by the synthesis routes developed for obtaining 2D-MXenes.
This variety of surface terminations enables a wealth of diverse compositions and stoichiometries with effects on the interaction between layers, and hence interlayer spacing, stacking properties and phonon scattering \cite{Hu_2016, Hadler_2021}, making  surface chemistry or intercalations a powerful tool for tuning thermal 
properties \cite{Gao_2020, Gogotsi_2025}. In this context, the fundamental heat transport mechanisms that lead to MXenes' intrinsic low thermal conductivities could be very diverse and remain unclear. These heat transfer mechanisms could include strong phonon scattering, interlayer and termination effects and electron-phonon coupling, among others.
%

Thermal measurements of MXenes have been carried out mainly on films made of a stack of Ti\textsubscript{3}C\textsubscript{2}T\textsubscript{\textit{x}} flakes, the most widely studied material in this family. The experimental results cover nearly an order of magnitude,\cite{Gogotsi_2025} from 0.14 to 0.8 W\,m\textsuperscript{$-1$}K\textsuperscript{$-1$}
in the out-of-plane (OOP) direction~\cite{Nguyen_2021,Wang_2022, Ouyang_2022, Kong_2023, Gehring_2024, Naqvi_2025}, a span too large to be due to measurement uncertainty alone. 
Similar variability is found in the in-plane (IP) direction \cite{Chen_2018,Liu_2023,Gehring_2024,Kumari_2025, Naqvi_2025}. 
Simulations, using both Molecular Dynamics (MD) and Density Functional Theory (DFT) methods \cite{Gholivand_2019, Wang_2022, Thanasarnsurapong_2025, Zhang_2026}, have been unable to resolve this discrepancy and consistently overestimate experimental values. This discrepancy has recently been singled out as an open problem in the field \cite{Gogotsi_2025}.
A frequently overlooked difference between experiments and simulations is that the latter model MXenes with homogeneous 
surface terminations \cite{Gholivand_2019}, while real samples contain heterogeneous surface terminations
\cite{Halim_2016, Naguib_2021}. We argue that this so-far unaccounted diversity of surface terminations leads 
to emerging effects that explain the discrepancy between measurements and simulations. 



Here, we present  a systematic Molecular Dynamics model of OOP phonon thermal conductivity in MXenes.
Using Non-Equilibrium Molecular Dynamics Simulations (NEMD), we explore the role that both the identity and relative content of surface terminations have on thermal conductivity in single 
Ti\textsubscript{3}C\textsubscript{2}T\textsubscript{\textit{x}} flakes. 
We find that even small deviations from homogeneous surface terminations drastically suppress thermal conductivity, bringing computed values into the range of experimental measurements. 
We also identify the mechanism responsible for this suppression: the simultaneous presence of surface terminations of different size generates a stereochemically induced vacuum gap between the layers of Ti\textsubscript{3}C\textsubscript{2}T\textsubscript{\textit{x}}, which leads to a low thermal conductivity.
By replacing surface terminations with larger chemical species, this gap can be deliberately widened, reducing the thermal conductivity to values an order of magnitude below those of homogeneous terminations, and below the minimum thermal conductivity limit for disordered solids \cite{Cahill_1992}.
Our results explain the long-standing variability in MXene thermal conductivity, while elucidating the relation between their surface chemistry and heat transport properties, thus providing a foundation for termination-intercalation strategies aimed at modulating the thermal conductivity.



\begin{figure*}[!ht]
\includegraphics[width=0.99\columnwidth]{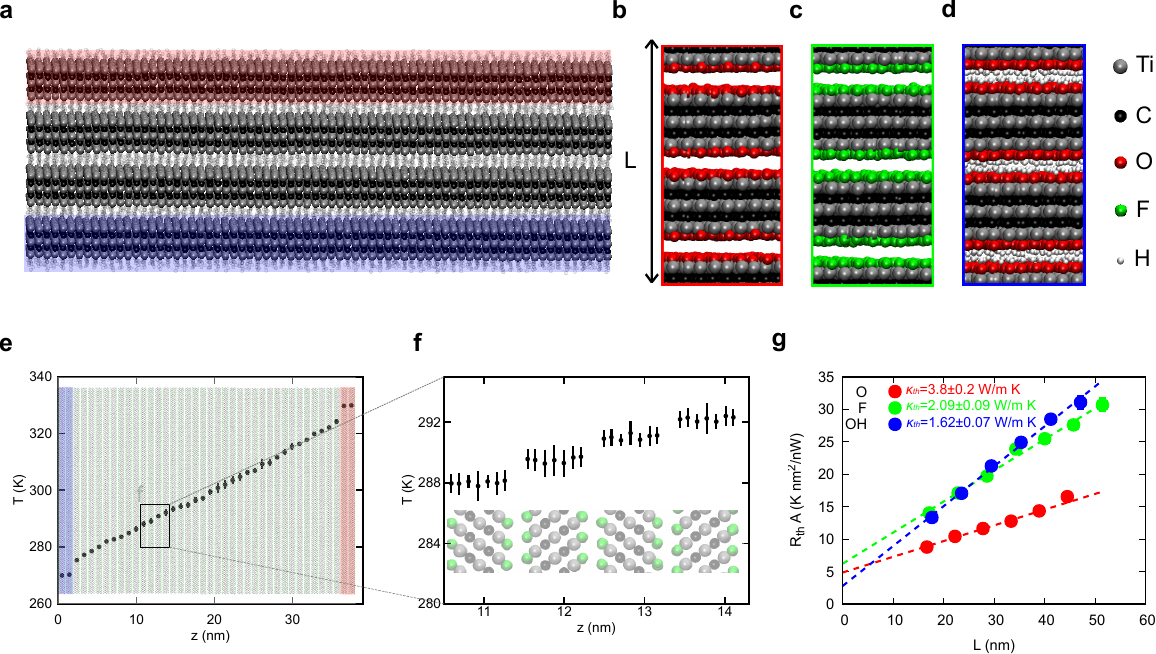}\\
\caption{
(a) Schematic of the simulation setup. A series of Ti\textsubscript{3}C\textsubscript{2}T\textsubscript{\textit{x}} are 
stacked, with extremes kept at different temperatures. A heat flux is allowed to form from the hot extreme to 
the cold one. (b-d) Schematics of Ti\textsubscript{3}C\textsubscript{2}T\textsubscript{\textit{x}} where the surface 
terminations are exclusively O (b), F (c) or OH (d). (e) Representative temperature profile in
Ti\textsubscript{3}C\textsubscript{2}F\textsubscript{2}. The temperature of each 
Ti\textsubscript{3}C\textsubscript{2}F\textsubscript{2} layer is represented as a function of its position. 
Note that the first and last two layers are connected to thermal baths. 
(f) Zoom of panel (e), showing independently the temperature in each Ti, C and F layer. 
(g) Thermal resistance times cross-section area as a function of system thickness for Ti\textsubscript{3}C\textsubscript{2}O\textsubscript{2},
Ti\textsubscript{3}C\textsubscript{2}F\textsubscript{2}, and Ti\textsubscript{3}C\textsubscript{2}(OH)\textsubscript{2}.
The dashed lines correspond to linear fits to the results and the extracted values are indicated in the legend.
}
\label{Fig_1}
\end{figure*}

In all simulations, we describe a single Ti\textsubscript{3}C\textsubscript{2}T\textsubscript{\textit{x}} flake, formed by a series of stacked layers, with flake thickness ranging between 15 nm and 50 nm.
The system has a cross-section area of 11.5 nm\textsuperscript{2}, large enough to prevent finite-size effects (Figure S1 and Figure S2).
The two outermost layers are held at different temperatures: 270 K and 330 K \cite{Brunger_1984}. 
This results in a temperature gradient in the OOP direction, which induces a heat flux (Figure~\ref{Fig_1}a). 
From the heat flux, we compute the phonon thermal conductivity, 
which is the dominant contribution to thermal transport due to the low electrical conductivity in the OOP direction 
\cite{Hu_2015, DeLeuze_2025}. 

We simulate O, F and OH groups as surface terminations, which are compatible with a synthesis route that includes etching with HF \cite{Naguib_2012, Naguib_2021}. In addition, to explore the influence of residual hydration that may persist after etching and incomplete drying, we additionally consider a hydrated OH/F surface motif denoted OH\textsubscript{2}F \cite{Celerier_2019, Zaman_2021}.

The interatomic potential is based on the INTERFACE force field (IFF) \cite{Heinz_2013} for MXenes, and was previously validated against MXene structural parameters, interlayer spacing, and vibrational spectra \cite{Armstrong_2025}, which are critical descriptors governing phonon-mediated heat transport. Recent work further demonstrated quantitative prediction of MXene interfacial and mechanical properties \cite{Armstrong_2025}. 
 
All simulations were performed in the LAMMPS software \cite{LAMMPS}. 
Further details on the force field and simulation protocol may be found in the Supporting Information. 
To understand how the heterogeneity of the surface terminations affects OOP thermal transport, we shall compare it with the idealized case of layers bearing a single termination species (Figure~\ref{Fig_1}~b--d).

Figure \ref{Fig_1}e shows a representative temperature profile of 
Ti\textsubscript{3}C\textsubscript{2}F\textsubscript{2}. 
Similar profiles for homogeneous O and OH surface terminations are shown in Figure S6. 
We see a linear dependence of the temperature on the 
position, suggesting an overall diffusive behavior.
When we zoom into the temperature within each layer of atoms (Figure \ref{Fig_1}f), we observe that the largest jumps in temperature, which correspond to main sources of thermal resistance, occur between the surface terminations of neighboring MXene layers. 
On the other hand, temperature within each MXene layer remains constant. 
As the surface terminations present more resistance to the heat flux than the titanium carbide structure, we can safely conclude that the thermal conduction will be dominated by the surface chemistry.

The thermal conductivity is computed in our simulations from values of thermal resistance as a function of the system thickness (Figure \ref{Fig_1}g), as shown in the Supporting Information (Figure S5). 
When we compare the computed thermal conductivity values for all three terminations, we observe a significant difference between them, 
ranging from 1.6~W\,m\textsuperscript{$-1$}K\textsuperscript{$-1$} in the case of OH surface terminations, to 3.8~W\,m\textsuperscript{$-1$}K\textsuperscript{$-1$} in the case of O. 
These results confirm that surface chemistry is the dominant effect in thermal conductivity, as we had anticipated from the temperature profiles. 
Yet when we compare these values to the range of experimental measurements, which lie between 0.14 and 0.8~W\,m\textsuperscript{$-1$}K\textsuperscript{$-1$}\cite{Nguyen_2021, Naqvi_2025, Gogotsi_2025},
we find that the simulations overestimate thermal conductivity. 
Previous studies on IP thermal conductivity show a similar trend, where calculations making use of both Density Functional Theory \cite{Gholivand_2019, Zhang_2026} and Molecular Dynamics simulations \cite{Wang_2022, Thanasarnsurapong_2025} overestimate experimental results 
\cite{Chen_2018, Naqvi_2025}.


The reason for this discrepancy relates to the fact that previous computational studies mainly focused on the analysis of the thermal conduction properties of systems with homogeneous terminations.
However, surface terminations are known to be mixed in realistic MXene samples \cite{Halim_2016, Naguib_2021}. 
Specifically, MXene surfaces frequently retain hydration species and residual water after etching, delamination and drying. The precise local arrangement of O, OH and F species remains difficult to determine experimentally and depends strongly on synthesis conditions, post-processing and environmental exposure \cite{Naguib_2021, Benchakar_2020, Seredych_2019}. 

\begin{figure}[!ht]
\includegraphics[width=0.99\columnwidth]{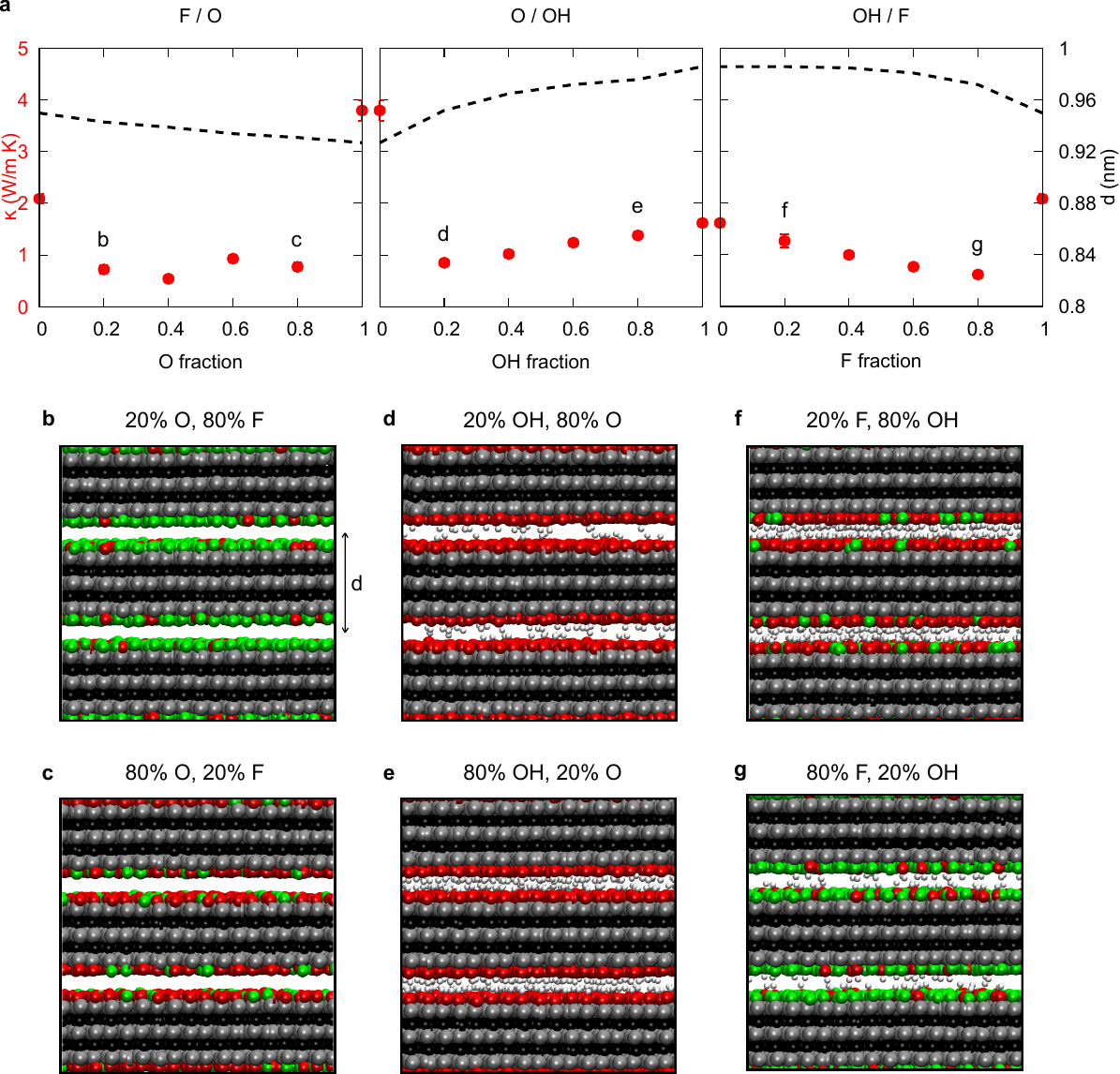}\\
\caption{
(a) Out-of-plane thermal conductivity in Ti\textsubscript{3}C\textsubscript{2}T\textsubscript{\textit{x}} (red dots)
for mixtures of F and O surface terminations (left), O and OH surface terminations (center), and OH and F surface terminations (right). The dashed lines correspond to the thickness of a 
Ti\textsubscript{3}C\textsubscript{2}T\textsubscript{\textit{x}} layer (see right vertical scale). 
(b-g) Selected snapshots from the NEMD simulations corresponding to different proportions of F and O surface 
terminations (b,c), O and OH terminations (d,e) and OH and F terminations (f,g).
}
\label{Fig_2}
\end{figure}

Such structural variability suggests that the local steric effects and interlayer spacing may vary substantially between similar samples. 
For this reason, we studied the thermal conductivity in systems with different surface terminations as a function of their relative proportion and the results are summarized in Figure~\ref{Fig_2}a. 
To be precise, we analyzed mixtures of two different terminations to better understand the influence of each surface termination. 
The corresponding data for thermal resistance as a function of system thickness may be found in the Supporting Information (Figure S7).
The most important conclusion of these simulations is that the simultaneous presence of two surface terminations drastically suppresses thermal conductivity to a range between 0.5 and
1.4~W\,m\textsuperscript{$-1$}K\textsuperscript{$-1$}. 
This range now overlaps with the experimental measurements.

Another important feature of the results of Figure~\ref{Fig_2}a is that the mixtures of F- and O- surface terminations yield the most insulating systems, yet there is no obvious correlation between thermal conductivity and the ratio between terminations. 
In contrast, both O- and OH- and F- and OH- mixtures present a common trend, 
namely both mixtures tend to exhibit higher thermal conductivity the higher the concentration of OH- surface terminations is.

To gain some insight into these results, we also present in Figure~\ref{Fig_2}a the thickness of a Ti\textsubscript{3}C\textsubscript{2}T\textsubscript{\textit{x}} layer $d$ as a function of the proportion of surface terminations.
We notice that the layer thickness values for systems with homogeneous terminations are 9.27 \text{\AA} for Ti\textsubscript{3}C\textsubscript{2}O\textsubscript{2}, 9.50 \text{\AA} for Ti\textsubscript{3}C\textsubscript{2}F\textsubscript{2}; and 9.86 \text{\AA} for  Ti\textsubscript{3}C\textsubscript{2}(OH)\textsubscript{2}.
Remarkably, in the case of heterogeneous terminations containing the bulkier OH-terminations, this thickness remains nearly constant at approximately 9.8 \text{\AA} as the OH- proportion decreases, and only changes significantly when its concentration vanishes. 
Previous results on the structure of  Ti\textsubscript{3}C\textsubscript{2}T\textsubscript{\textit{x}}, from X-Ray Diffraction (XRD) \cite{Mashtalir_2013, Ferrara_2021} and Scanning Electron Microscopy (SEM) \cite{Wang_2015, Dong_2017, Ferrara_2021}, also estimated the layer thickness in Ti\textsubscript{3}C\textsubscript{2}T\textsubscript{\textit{x}} as 9.8 \text{\AA}, which we theorize is due to the presence of -OH surface terminations after synthesis, as shown by X-Ray photoelectron spectroscopy \cite{Halim_2016, Benchakar_2020} and thermogravimetric analysis \cite{Seredych_2019}.

We attribute this observation to the fact that the presence of OH- surface terminations, bulkier than F- or O- terminations, opens a larger gap between Ti\textsubscript{3}C\textsubscript{2}T\textsubscript{\textit{x}} layers. Another important piece of information is revealed by the snapshots from NEMD simulations shown in Figure~\ref{Fig_2}(b--g). 
They show that in systems with smaller proportions of OH terminations, which are more insulating, the interlayer space is emptier than in more conducting systems.

These different observations strongly suggest that the main reason for the low thermal conductivity observed in OOP Ti\textsubscript{3}C\textsubscript{2}T\textsubscript{\textit{x}} is the stereochemically induced gap formed by the presence of surface terminations of different size. 
This is the central idea of this work and it naturally explains the 
different conductivity trends seen in Figure \ref{Fig_2}a. 
First, a small proportion of OH- surface terminations opens a large gap between layers, which drastically reduces the thermal conductivity as compared to homogeneous systems. 
On the other hand, as the content in OH- terminations increases, this interlayer gap is filled by atoms, which results in more conductive channels and an increase in thermal conductivity. 
We show in the Supporting Information (Figure S8) that this trend holds when all three surface terminations (O, F and OH) are present as well.

\begin{figure}[!hb]
\includegraphics[width=0.60\columnwidth]{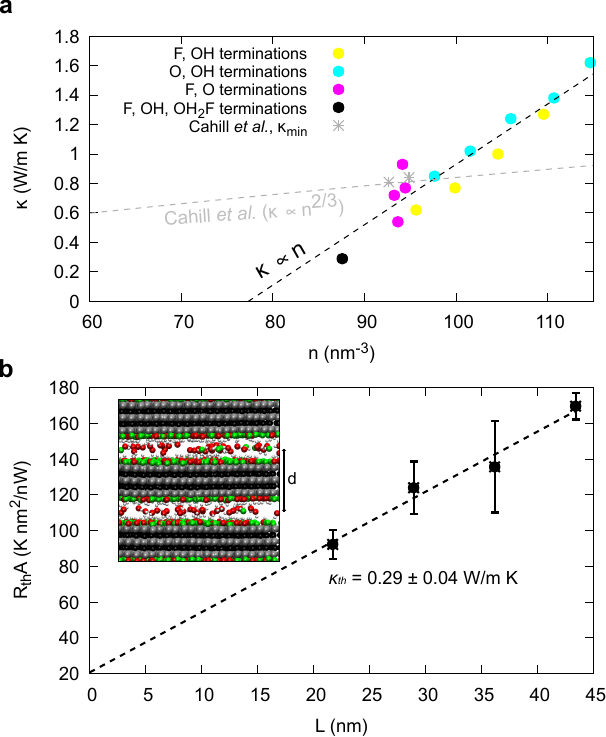}\\
\caption{
(a) Thermal conductivity values shown in Figure \ref{Fig_2}a as a function of the number density defined in
\eqref{Num_den}. Values in magenta correspond to F and O terminations, values in cyan correspond to O and OH 
terminations, and values in yellow correspond to OH and F terminations. 
The value in black corresponds to the system shown in panel (b). 
Values in grey correspond to the minimum thermal conductivity model \cite{Cahill_1992} for Ti\textsubscript{3}C\textsubscript{2}F\textsubscript{2} and Ti\textsubscript{3}C\textsubscript{2}O\textsubscript{2}, as described in the Supporting Information (Figure S9).
The line in grey represents thermal conductivity as predicted by the minimum thermal conductivity model \cite{Cahill_1992}, $\kappa_{\min} \propto n^{2/3}$, computed through the finite displacement method \cite{Togo_2015}, while the line in black represents a linear fit, $\kappa \propto n$, of thermal conductivity values computed from Molecular Dynamics simulations, as a comparison.
(b) Out-of-plane thermal resistance times cross-section area as 
a function of system thickness for Ti\textsubscript{3}C\textsubscript{2}T\textsubscript{\textit{x}}. Surface terminations correspond to $50\%$ F, $40\%$ OH and $10\%$ OH\textsubscript{2}F.
}
\label{Fig_3}
\end{figure}

 
The previous arguments lead to the natural conclusion that the main parameter determining the thermal conductivity in MXenes is the number of atoms in the interlayer space. 
To express this idea in quantitative terms, we define the atom number density as the number of atoms in a layer per unit of volume:
\begin{equation}
    n = \frac{N}{V}= \frac{N}{Ad},
    \label{Num_den}
\end{equation}
where $N$ is the number of atoms in a Ti\textsubscript{3}C\textsubscript{2}T\textsubscript{\textit{x}} layer, $A$ is the cross-section area, and $d$ is the layer thickness. Since the Ti\textsubscript{3}C\textsubscript{2} core is identical in all our systems, variations in $n$ directly reflect the termination content and the interlayer spacing.
 
Thermal conductivity is expected to depend on this atom number density. Indeed, Cahill \textit{et al.} \cite{Cahill_1992} developed a successful model for the minimum thermal conductivity of a solid, in which this conductivity was proportional to $n^{2/3}$. However, the wide dispersion of thermal conductivity values shown in Figure \ref{Fig_2} seems to indicate a stronger dependence on the atom density, pointing to the stereochemical gap as a strong mechanism to suppress thermal conductivity. Indeed, our computed values follow a markedly steeper scaling with the atomic density than the $n^{2/3}$ dependence of the model. This suggests that the stereochemical gap suppresses heat transport more strongly than the overdamped, maximally scattered transport the model assumes, although a systematic study of other layered materials is called for to establish the generality of this scaling.
 
In Figure~\ref{Fig_3}a we present the ensemble of conductivity results of Figure~\ref{Fig_2}a, as a function of the atom number density. 
We observe a strong correlation between conductivity and density, showing that thermally insulating MXenes are achieved by a small proportion of OH terminations, allowing a large, empty stereochemically induced gap in the space between layers. 
From Figure \ref{Fig_3}a we can also conclude that the variability in thermal conductivity values ultimately stems from different proportions of OH- surface terminations, and therefore different atom densities, across different systems.
 
The relation between atom density and thermal conductivity points towards a general strategy to tune thermal conductivity in MXenes. 
A surface chemistry that minimizes this density, for example, with a small concentration of larger species attached to the surface, would further reduce the atom density, leading to more insulating systems. Given the strong dependence between thermal conductivity and density shown in Figure \ref{Fig_3}, we propose that tuning of the stereochemical gap may allow  MXenes to reach thermal conductivities lower than the predictions of the minimum thermal conductivity model \cite{Cahill_1992}, as it has been reported for other stacked 2D materials such as WSe\textsubscript{2} \cite{Cahill_2007}, MoS\textsubscript{2} or WS\textsubscript{2}\cite{Erhart_Cahill_Park_2021}.
 
To test this idea, we analyzed a system in which part of the surface terminations are replaced by a larger species, allowing us to open a larger stereochemical gap.
Motivated by experimental observations that residual water and hydrated surface species may remain associated with O-, OH- and F-terminated MXene surfaces after etching and drying \cite{Celerier_2019, Zaman_2021}, we explored the effect of a larger hydrated surface motif.
Specifically, 10\% of the surface terminations were replaced by a model OH\textsubscript{2}F species while maintaining the same elemental composition already present on MXene surfaces.
Our results, shown in Figure \ref{Fig_3}b, indicate that, in this composition, thermal conductivity is reduced to near 0.3 W\,m\textsuperscript{$-1$}K\textsuperscript{$-1$}, well within the range of experimental values, and an order of magnitude less 
than values computed for systems with homogeneous terminations. This value lies below both the minimum thermal conductivity computed for the homogeneous phases and the $n^{2/3}$ extrapolation of the model at the corresponding atom density (Figure \ref{Fig_3}a). Taken together with the lowest experimental reports, which also fall below this limit, these results establish the stereochemical gap as a route to sub-minimum out-of-plane heat transport in MXenes.
 
 
 
To summarize, in this work we have presented a theoretical study of the thermal conductivity in the out-of-plane direction in Ti\textsubscript{3}C\textsubscript{2}T\textsubscript{\textit{x}} for systems with both homogeneous and heterogeneous surface terminations. 
We have shown that the highly insulating properties of OOP Ti\textsubscript{3}C\textsubscript{2}T\textsubscript{\textit{x}} emerge from the presence of a stereochemically induced gap, formed by the simultaneous presence of surface terminations with different sizes. 
This naturally explains the discrepancies reported thus far in the literature between experiments and simulations. 
We have also demonstrated that the value of thermal conductivity can be related to the relative content of OH surface terminations filling this gap. Widening this gap with bulkier hydrated species drives the out-of-plane thermal conductivity below the minimum thermal conductivity limit \cite{Cahill_1992}, placing MXenes, together with WSe\textsubscript{2} \cite{Cahill_2007} and MoS\textsubscript{2}/WS\textsubscript{2} \cite{Erhart_Cahill_Park_2021}, among the layered materials known to exhibit sub-minimum out-of-plane heat transport.
These results further suggest that residual hydration and drying history may contribute to the substantial variability of experimentally reported thermal conductivities in Ti\textsubscript{3}C\textsubscript{2}T\textsubscript{\textit{x}} films, since even small populations of larger hydrated species can significantly alter the effective stereochemically induced gap.
 
Our work elucidates the effect that surface chemistry has on the thermal conductivity of MXenes, showing this effect is more structural than chemical in origin. 
Therefore, while the surface chemistry composition of MXenes is often impossible to predict \cite{Naguib_2021}, the results presented here offer a solid foundation for thermal engineering strategies \cite{Gao_2020, Gogotsi_2025} that modulate the heat transport across MXenes by controlling the intercalated surface terminations or species in the interlayer.




%
%
\begin{suppinfo}
The Supporting Information document is available free of charge at --- and includes the following: 
\begin{itemize}
 \item S1. Computational methods.   
 \item S2. Temperature profiles for Ti$_{3}$C$_{2}$T$_{x}$.
 \item S3. Thermal resistance in inhomogeneous terminations.
 \item S4. Thermal conductivity in mixed terminations.
 \item S5. Minimum thermal conductivity model.
 \end{itemize}
\end{suppinfo}

\section*{Author Information}

\subsection*{Corresponding Authors}
*E-mail: juancarlos.cuevas@uam.es (J.C.C.);
*E-mail: m.m.rojo@csic.es (M.M.R.).

\subsection*{Author Contributions}
J.G.V., M.M.R., and J.C.C. conceived the work;
O.M.L. conducted all the numerical simulations and analyzed the results with the help of
J.G.V. and J.C.C.;
I.A. and H.H. developed the force field;
O.M.L., J.G.V., and J.C.C. wrote the manuscript with input from all authors.

\subsection*{Notes}
The authors declare no competing financial interest.

\begin{acknowledgement}
We acknowledge support from the Spanish Ministry of Science, Innovation and Universities \& the State Research 
Agency MICIU/ AEI/10.13039/501100011033, through the grant numbers PID2020-113722RJ-I00,  CNS2023-144011, 
CEX2023-001316-M, CEX2024-001445-S, PID2024-157536NB-C21, PID2024-157536NB-C22, “María de Maeztu” Programme 
for Units of Excellence in R\&D (CEX2023- 001316-M), the Severo Ochoa Centres of Excellence program 
(CEX2024-001445-S) and the Spanish CM “Talento Program” Project No.\ 2020-T1/ND-20306.
This work utilized research computing resources at Picasso, Finisterrae3 (award number RES-FI-2025-2-0056). We also acknowledge the support from the European Union (ERC CoG THERMO2DEAL, 101123381). Views and opinions expressed are however those of the authors only and do not necessarily reflect those of the European Union or the European Research Council. Neither the European Union nor the granting authority can be held responsible for them.
\end{acknowledgement}


\cleardoublepage

\bibliography{Bibliography}

@article{Gehring_2024,
        title = {Violation of the {Wiedemann}–{Franz} {Law} and {Ultralow} {Thermal} {Conductivity} of {Ti}$_{\textrm{3}}$ {C}$_{\textrm{2}}$ {T}$_{\textrm{ \textit{x} }}$ {MXene}},
        volume = {18},
        pages = {32491--32497},
        year = {2024},
        journal = {ACS Nano},
        author = {Huang, Yubin and Spiece, Jean and Parker, Tetiana and Lee, Asaph and Gogotsi, Yury and Gehring, Pascal},
        copyright = {https://doi.org/10.15223/policy-029},
        issn = {1936-0851, 1936-086X},
        url = {https://pubs.acs.org/doi/10.1021/acsnano.4c08189},
        doi = {10.1021/acsnano.4c08189},
}

@article{Erhart_Cahill_Park_2021,
        title = {Extremely anisotropic van der {Waals} thermal conductors},
        volume = {597},
        issn = {0028-0836, 1476-4687},
        url = {https://www.nature.com/articles/s41586-021-03867-8},
        doi = {10.1038/s41586-021-03867-8},
        language = {en},
        number = {7878},
        urldate = {2026-03-24},
        journal = {Nature},
        author = {Kim, Shi En and Mujid, Fauzia and Rai, Akash and Eriksson, Fredrik and Suh, Joonki and Poddar, Preeti and Ray, Ariana and Park, Chibeom and Fransson, Erik and Zhong, Yu and Muller, David A. and Erhart, Paul and Cahill, David G. and Park, Jiwoong},
        month = sep,
        year = {2021},
        pages = {660--665},
}

@article{Cahill_2007,
        title = {Ultralow {Thermal} {Conductivity} in {Disordered}, {Layered} {WSe}$_{\textrm{2}}$ {Crystals}},
        volume = {315},
        issn = {0036-8075, 1095-9203},
        url = {https://www.science.org/doi/10.1126/science.1136494},
        doi = {10.1126/science.1136494},
        language = {en},
        number = {5810},
        urldate = {2026-03-24},
        journal = {Science},
        author = {Chiritescu, Catalin and Cahill, David G. and Nguyen, Ngoc and Johnson, David and Bodapati, Arun and Keblinski, Pawel and Zschack, Paul},
        month = jan,
        year = {2007},
        pages = {351--353},
}

@article{Naguib_2021,
        title = {Ten {Years} of {Progress} in the {Synthesis} and {Development} of {MXenes}},
        volume = {33},
        issn = {0935-9648, 1521-4095},
        url = {https://advanced.onlinelibrary.wiley.com/doi/10.1002/adma.202103393},
        doi = {10.1002/adma.202103393},
        language = {en},
        number = {39},
        urldate = {2026-02-19},
        journal = {Advanced Materials},
        author = {Naguib, Michael and Barsoum, Michel W. and Gogotsi, Yury},
        month = oct,
        year = {2021},
        pages = {2103393},
}

@article{Gogotsi_2023,
        title = {The {Future} of {MXenes}},
        volume = {35},
        copyright = {https://doi.org/10.15223/policy-001},
        issn = {0897-4756, 1520-5002},
        url = {https://pubs.acs.org/doi/10.1021/acs.chemmater.3c02491},
        doi = {10.1021/acs.chemmater.3c02491},
        language = {en},
        number = {21},
        urldate = {2026-02-19},
        journal = {Chemistry of Materials},
        author = {Gogotsi, Yury},
        month = nov,
        year = {2023},
        pages = {8767--8770},
}

@article{Lipatov_2020,
        title = {Electrical and {Elastic} {Properties} of {Individual} {Single}‐{Layer} {Nb}$_{\textrm{4}}$ {C}$_{\textrm{3}}$ {T} \textit{$_{\textrm{x}}$ } {MXene} {Flakes}},
        volume = {6},
        issn = {2199-160X, 2199-160X},
        url = {https://advanced.onlinelibrary.wiley.com/doi/10.1002/aelm.201901382},
        doi = {10.1002/aelm.201901382},
        language = {en},
        number = {4},
        urldate = {2026-03-24},
        journal = {Advanced Electronic Materials},
        author = {Lipatov, Alexey and Alhabeb, Mohamed and Lu, Haidong and Zhao, Shuangshuang and Loes, Michael J. and Vorobeva, Nataliia S. and Dall'Agnese, Yohan and Gao, Yu and Gruverman, Alexei and Gogotsi, Yury and Sinitskii, Alexander},
        month = apr,
        year = {2020},
        pages = {1901382},
}

@article{Naguib_2011,
	title = {Two‐{Dimensional} {Nanocrystals} {Produced} by {Exfoliation} of {Ti}$_{\textrm{3}}$ {AlC}$_{\textrm{2}}$},
	volume = {23},
	copyright = {http://onlinelibrary.wiley.com/termsAndConditions\#vor},
	issn = {0935-9648, 1521-4095},
	url = {https://advanced.onlinelibrary.wiley.com/doi/10.1002/adma.201102306},
	doi = {10.1002/adma.201102306},
	language = {en},
	number = {37},
	urldate = {2026-02-18},
	journal = {Advanced Materials},
	author = {Naguib, Michael and Kurtoglu, Murat and Presser, Volker and Lu, Jun and Niu, Junjie and Heon, Min and Hultman, Lars and Gogotsi, Yury and Barsoum, Michel W.},
	month = oct,
	year = {2011},
	pages = {4248--4253},
}

@article{Naguib_2012,
	title = {Two-{Dimensional} {Transition} {Metal} {Carbides}},
	volume = {6},
	issn = {1936-0851, 1936-086X},
	url = {https://pubs.acs.org/doi/10.1021/nn204153h},
	doi = {10.1021/nn204153h},
	language = {en},
	number = {2},
	urldate = {2026-02-19},
	journal = {ACS Nano},
	author = {Naguib, Michael and Mashtalir, Olha and Carle, Joshua and Presser, Volker and Lu, Jun and Hultman, Lars and Gogotsi, Yury and Barsoum, Michel W.},
	month = feb,
	year = {2012},
	pages = {1322--1331},
}

@article{Naguib_2014,
	title = {25th {Anniversary} {Article}: {MXenes}: {A} {New} {Family} of {Two}‐{Dimensional} {Materials}},
	volume = {26},
	copyright = {http://onlinelibrary.wiley.com/termsAndConditions\#vor},
	issn = {0935-9648, 1521-4095},
	shorttitle = {25th {Anniversary} {Article}},
	url = {https://advanced.onlinelibrary.wiley.com/doi/10.1002/adma.201304138},
	doi = {10.1002/adma.201304138},
	language = {en},
	number = {7},
	urldate = {2026-02-19},
	journal = {Advanced Materials},
	author = {Naguib, Michael and Mochalin, Vadym N. and Barsoum, Michel W. and Gogotsi, Yury},
	month = feb,
	year = {2014},
	pages = {992--1005},
}

@article{Anasori_2017,
        title = {{2D} metal carbides and nitrides ({MXenes}) for energy storage},
        language = {en},
        journal = {Nature Review Materials},
        author = {Anasori, Babak and Lukatskaya, Maria R and Gogotsi, Yury},
        year = {2017},
}

@article{Pang_2019,
        title = {Applications of {2D} {MXenes} in energy conversion and storage systems},
        volume = {48},
        issn = {0306-0012, 1460-4744},
        url = {https://xlink.rsc.org/?DOI=C8CS00324F},
        doi = {10.1039/C8CS00324F},
        language = {en},
        number = {1},
        urldate = {2026-02-19},
        journal = {Chemical Society Reviews},
        author = {Pang, Jinbo and Mendes, Rafael G. and Bachmatiuk, Alicja and Zhao, Liang and Ta, Huy Q. and Gemming, Thomas and Liu, Hong and Liu, Zhongfan and Rummeli, Mark H.},
        year = {2019},
        pages = {72--133},
}

@article{Naguib_2013,
        title = {New {Two}-{Dimensional} {Niobium} and {Vanadium} {Carbides} as {Promising} {Materials} for {Li}-{Ion} {Batteries}},
        volume = {135},
        issn = {0002-7863, 1520-5126},
        url = {https://pubs.acs.org/doi/10.1021/ja405735d},
        doi = {10.1021/ja405735d},
        language = {en},
        number = {43},
        urldate = {2026-02-19},
        journal = {Journal of the American Chemical Society},
        author = {Naguib, Michael and Halim, Joseph and Lu, Jun and Cook, Kevin M. and Hultman, Lars and Gogotsi, Yury and Barsoum, Michel W.},
        month = oct,
        year = {2013},
        pages = {15966--15969},
}

@article{Halim_2014,
        title = {Transparent {Conductive} {Two}-{Dimensional} {Titanium} {Carbide} {Epitaxial} {Thin} {Films}},
        volume = {26},
        copyright = {http://pubs.acs.org/page/policy/authorchoice\_ccby\_termsofuse.html},
        issn = {0897-4756, 1520-5002},
        url = {https://pubs.acs.org/doi/10.1021/cm500641a},
        doi = {10.1021/cm500641a},
        language = {en},
        number = {7},
        urldate = {2026-02-19},
        journal = {Chemistry of Materials},
        author = {Halim, Joseph and Lukatskaya, Maria R. and Cook, Kevin M. and Lu, Jun and Smith, Cole R. and Näslund, Lars and May, Steven J. and Hultman, Lars and Gogotsi, Yury and Eklund, Per and Barsoum, Michel W.},
        month = apr,
        year = {2014},
        pages = {2374--2381},
}

@article{Halim_2016,
        title = {X-ray photoelectron spectroscopy of select multi-layered transition metal carbides ({MXenes})},
        volume = {362},
        issn = {01694332},
        url = {https://linkinghub.elsevier.com/retrieve/pii/S0169433215027841},
        doi = {10.1016/j.apsusc.2015.11.089},
        language = {en},
        urldate = {2026-01-16},
        journal = {Applied Surface Science},
        author = {Halim, Joseph and Cook, Kevin M. and Naguib, Michael and Eklund, Per and Gogotsi, Yury and Rosen, Johanna and Barsoum, Michel W.},
        month = jan,
        year = {2016},
        pages = {406--417},
}

@article{Hu_2016,
        title = {Interlayer coupling in two-dimensional titanium carbide {MXenes}},
        volume = {18},
        issn = {1463-9076, 1463-9084},
        url = {https://xlink.rsc.org/?DOI=C6CP01699E},
        doi = {10.1039/C6CP01699E},
        language = {en},
        number = {30},
        urldate = {2025-10-01},
        journal = {Physical Chemistry Chemical Physics},
        author = {Hu, Tao and Hu, Minmin and Li, Zhaojin and Zhang, Hui and Zhang, Chao and Wang, Jingyang and Wang, Xiaohui},
        year = {2016},
        pages = {20256--20260},
}

@article{Hadler_2021,
        title = {Stacking {Sequence}, {Interlayer} {Bonding}, {Termination} {Group} {Stability} and {Li}/{Na}/{Mg} {Diffusion} in {MXenes}},
        volume = {3},
        copyright = {https://creativecommons.org/licenses/by/4.0/},
        issn = {2639-4979, 2639-4979},
        url = {https://pubs.acs.org/doi/10.1021/acsmaterialslett.1c00316},
        doi = {10.1021/acsmaterialslett.1c00316},
        language = {en},
        number = {9},
        urldate = {2025-10-01},
        journal = {ACS Materials Letters},
        author = {Hadler-Jacobsen, Jacob and Fagerli, Frode Håskjold and Kaland, Henning and Schnell, Sondre Kvalvåg},
        month = sep,
        year = {2021},
        pages = {1369--1376},
}

@article{Gao_2020,
        title = {Tracking ion intercalation into layered {Ti}$_{\textrm{3}}$ {C}$_{\textrm{2}}$ {MXene} films across length scales},
        volume = {13},
        issn = {1754-5692, 1754-5706},
        url = {https://xlink.rsc.org/?DOI=D0EE01580F},
        doi = {10.1039/D0EE01580F},
        language = {en},
        number = {8},
        urldate = {2026-02-19},
        journal = {Energy \& Environmental Science},
        author = {Gao, Qiang and Sun, Weiwei and Ilani-Kashkouli, Poorandokht and Tselev, Alexander and Kent, Paul R. C. and Kabengi, Nadine and Naguib, Michael and Alhabeb, Mohamed and Tsai, Wan-Yu and Baddorf, Arthur P. and Huang, Jingsong and Jesse, Stephen and Gogotsi, Yury and Balke, Nina},
        year = {2020},
        pages = {2549--2558},
}

@article{Nguyen_2021,
        title = {Drastically increased electrical and thermal conductivities of {Pt}-infiltrated {MXenes}},
        volume = {9},
        issn = {2050-7488, 2050-7496},
        url = {https://xlink.rsc.org/?DOI=D1TA00331C},
        doi = {10.1039/D1TA00331C},
        language = {en},
        number = {17},
        urldate = {2026-03-10},
        journal = {Journal of Materials Chemistry A},
        author = {Nguyen, Viet Phuong and Lim, Mikyung and Kim, Kyung-Shik and Kim, Jae-Hyun and Park, Ji Su and Yuk, Jong Min and Lee, Seung-Mo},
        year = {2021},
        pages = {10739--10746},
}

@article{Ouyang_2022,
        title = {Synergistical thermal modulation function of {2D} {Ti3C2} {MXene} composite nanosheets via interfacial structure modification},
        volume = {25},
        issn = {25890042},
        url = {https://linkinghub.elsevier.com/retrieve/pii/S2589004222010975},
        doi = {10.1016/j.isci.2022.104825},
        language = {en},
        number = {8},
        urldate = {2026-03-24},
        journal = {iScience},
        author = {Ouyang, Yuxin and Qiu, Lin and Bai, Yangyang and Yu, Wei and Feng, Yanhui},
        month = aug,
        year = {2022},
        pages = {104825},
}

@article{Kong_2023,
        title = {Mannitol enhanced thermal conductivity and environmental stability of highly aligned {MXene} composite film},
        volume = {241},
        issn = {02663538},
        url = {https://linkinghub.elsevier.com/retrieve/pii/S0266353823002348},
        doi = {10.1016/j.compscitech.2023.110141},
        language = {en},
        urldate = {2026-03-24},
        journal = {Composites Science and Technology},
        author = {Kong, Xiangdong and Song, Guichen and Chen, Yapeng and Chen, Xuemei and Li, Maohua and Li, Linhong and Wang, Yandong and Gong, Ping and Zhang, Zhenbang and Zhang, Jianxiang and Yang, Rongjie and Xu, Kang and Cai, Tao and Chang, Keke and Pan, Zhongbin and Wang, Bo and Wu, Xinfeng and Lin, Cheng-Te and Nishimura, Kazuhito and Jiang, Nan and Yu, Jinhong},
        month = aug,
        year = {2023},
        pages = {110141},
}

@article{Wang_2022,
        title = {Thermal conductivities of {Ti3C2Tx} {MXenes} and their interfacial thermal performance in {MXene}/epoxy composites – a molecular dynamics simulation},
        volume = {194},
        issn = {00179310},
        url = {https://linkinghub.elsevier.com/retrieve/pii/S0017931022005002},
        doi = {10.1016/j.ijheatmasstransfer.2022.123027},
        language = {en},
        urldate = {2026-03-24},
        journal = {International Journal of Heat and Mass Transfer},
        author = {Wang, Menglin and Liu, Yifang and Zhang, Haoran and Wu, Yanbing and Pan, Lei},
        month = sep,
        year = {2022},
        pages = {123027},
}

@article{Naqvi_2025,
        title = {Surface {Functionalization} of {Ti}$_{\textrm{3}}$ {C}$_{\textrm{2}}$ {T}$_{\textrm{ \textit{x} }}$ {MXenes} in {Epoxy} {Nanocomposites}: {Enhancing} {Conductivity}, {EMI} {Shielding}, {Thermal} {Conductivity}, and {Mechanical} {Strength}},
        volume = {17},
        copyright = {https://doi.org/10.15223/policy-029},
        issn = {1944-8244, 1944-8252},
        shorttitle = {Surface {Functionalization} of {Ti}$_{\textrm{3}}$ {C}$_{\textrm{2}}$ {T}$_{\textrm{ \textit{x} }}$ {MXenes} in {Epoxy} {Nanocomposites}},
        url = {https://pubs.acs.org/doi/10.1021/acsami.4c21997},
        doi = {10.1021/acsami.4c21997},
        language = {en},
        number = {13},
        urldate = {2026-03-10},
        journal = {ACS Applied Materials \& Interfaces},
        author = {Naqvi, Shabbir Madad and Hassan, Tufail and Iqbal, Aamir and Jung, Sungmin and Jeong, Seunghwan and Zaman, Shakir and Zafar, Ujala and Hussain, Noushad and Cho, Sooyeong and Koo, Chong Min},
        month = apr,
        year = {2025},
        pages = {20149--20161},
}

@article{Kumari_2025,
        title = {Probing phonon anharmonicity of multilayer {T} i 3 {C} 2 {T} x ( {T} : {OH}, {O}, or {F}) {MXene} through temperature- and pressure-dependent {Raman} studies},
        volume = {111},
        issn = {2469-9950, 2469-9969},
        shorttitle = {Probing phonon anharmonicity of multilayer {T} i 3 {C} 2 {T} x ( {T}},
        url = {https://link.aps.org/doi/10.1103/PhysRevB.111.115411},
        doi = {10.1103/PhysRevB.111.115411},
        language = {en},
        number = {11},
        urldate = {2026-03-24},
        journal = {Physical Review B},
        author = {Kumari, Kaushalya and Thiyagarajan, R. and Kale, Abhijeet J. and Batra, Rohit and Krishanmurty, Srini and Sethupathi, K. and Rao, M. S. Ramachandra},
        month = mar,
        year = {2025},
        pages = {115411},
}

@article{Liu_2023,
        title = {Electrically insulating {PBO}/{MXene} film with superior thermal conductivity, mechanical properties, thermal stability, and flame retardancy},
        volume = {14},
        issn = {2041-1723},
        url = {https://www.nature.com/articles/s41467-023-40707-x},
        doi = {10.1038/s41467-023-40707-x},
        language = {en},
        number = {1},
        urldate = {2026-03-24},
        journal = {Nature Communications},
        author = {Liu, Yong and Zou, Weizhi and Zhao, Ning and Xu, Jian},
        month = sep,
        year = {2023},
        pages = {5342},
}

@article{Chen_2018,
        title = {Measurement and {Analysis} of {Thermal} {Conductivity} of {Ti3C2Tx} {MXene} {Films}},
        volume = {11},
        issn = {1996-1944},
        url = {https://www.mdpi.com/1996-1944/11/9/1701},
        doi = {10.3390/ma11091701},
        language = {en},
        number = {9},
        urldate = {2026-03-10},
        journal = {Materials},
        author = {Chen, Lin and Shi, Xuguo and Yu, Nanjie and Zhang, Xing and Du, Xiaoze and Lin, Jun},
        month = sep,
        year = {2018},
        pages = {1701},
}

@article{Thanasarnsurapong_2025,
        title = {Accelerating {Lattice} {Thermal} {Conductivity} {Calculations} in {MXenes}: {A} {Machine} {Learning} {Force} {Field} {Approach}},
        volume = {5},
        copyright = {https://creativecommons.org/licenses/by-nc-nd/4.0/},
        issn = {2694-2461, 2694-2461},
        shorttitle = {Accelerating {Lattice} {Thermal} {Conductivity} {Calculations} in {MXenes}},
        url = {https://pubs.acs.org/doi/10.1021/acsmaterialsau.5c00043},
        doi = {10.1021/acsmaterialsau.5c00043},
        language = {en},
        number = {5},
        urldate = {2025-09-29},
        journal = {ACS Materials Au},
        author = {Thanasarnsurapong, Thanasee and Jana, Sourav Kanti and Detrattanawichai, Panyalak and Namunmong, Waraporn and Hirunpinyopas, Wisit and Iamprasertkun, Pawin and Boonchun, Adisak},
        month = sep,
        year = {2025},
        pages = {823--830},
}

@article{Gholivand_2019,
        title = {Effect of surface termination on the lattice thermal conductivity of monolayer {Ti3C2Tz} {MXenes}},
        volume = {126},
        issn = {0021-8979, 1089-7550},
        url = {https://pubs.aip.org/jap/article/126/6/065101/156797/Effect-of-surface-termination-on-the-lattice},
        doi = {10.1063/1.5094294},
        language = {en},
        number = {6},
        urldate = {2026-02-19},
        journal = {Journal of Applied Physics},
        author = {Gholivand, Hamed and Fuladi, Shadi and Hemmat, Zahra and Salehi-Khojin, Amin and Khalili-Araghi, Fatemeh},
        month = aug,
        year = {2019},
        pages = {065101},
}

@article{Zhang_2026,
        title = {Thermal transport in {Ti3C2} {MXene} enhanced by hydrogen functionalization-induced symmetry reconstruction},
        volume = {139},
        issn = {0021-8979, 1089-7550},
        url = {https://pubs.aip.org/jap/article/139/6/065103/3379989/Thermal-transport-in-Ti3C2-MXene-enhanced-by},
        doi = {10.1063/5.0310379},
        language = {en},
        number = {6},
        urldate = {2026-02-19},
        journal = {Journal of Applied Physics},
        author = {Zhang, Guangwu and Cheng, Xue and Yang, Chao and Shao, Cheng and Wang, Xinyu},
        month = feb,
        year = {2026},
        pages = {065103},
}

@article{Gogotsi_2025,
	title = {{MXenes} for {Infrared} {Thermal} {Management}},
	volume = {19},
	copyright = {https://creativecommons.org/licenses/by/4.0/},
	issn = {1936-0851, 1936-086X},
	url = {https://pubs.acs.org/doi/10.1021/acsnano.5c14464},
	doi = {10.1021/acsnano.5c14464},
	language = {en},
	number = {48},
	urldate = {2026-02-04},
	journal = {ACS Nano},
	author = {Hassan, Tufail and Park, Changhoon and Naqvi, Shabbir Madad and Kim, Jongyoun and Kim, Hyunho and Khalid, Zubair and Jung, Sungmin and Gogotsi, Yury and Koo, Chong Min},
	month = dec,
	year = {2025},
	pages = {40703--40732},
}

@article{Brunger_1984,
        title = {Stochastic Boundary Conditions for Molecular Dynamics Simulations of ST2 Water},
        volume = {105},
        language = {en},
        number = {5},
        journal = {Chemical Physics Letters},
        author = {Brunger, Axel and Brooks, Charles and Karplus, Martin},
        year = {1984},
}

@article{LAMMPS,
  author = {A. P. Thompson and H. M. Aktulga and R. Berger and 
     D. S. Bolintineanu and W. M. Brown and P. S. Crozier and
     P. J. in 't Veld and A. Kohlmeyer and S. G. Moore and T. D. Nguyen and
     R. Shan and M. J. Stevens and J. Tranchida and C. Trott and S. J. Plimpton},
  title = {{LAMMPS} - a flexible simulation tool for
     particle-based materials modeling at the 
     atomic, meso, and continuum scales},
  journal = {Comp. Phys. Comm.},
  volume =  {271},
  pages =   {108171},
  year =    {2022},
  doi = {10.1016/j.cpc.2021.108171}
}

@article{DeLeuze_2025,
        title = {Anisotropic charge transport in {2D} single crystals of {Ti3C2Tx} {MXenes}},
        volume = {6},
        issn = {2662-4443},
        url = {https://www.nature.com/articles/s43246-025-00902-3},
        doi = {10.1038/s43246-025-00902-3},
        language = {en},
        number = {1},
        urldate = {2026-03-19},
        journal = {Communications Materials},
        author = {De Leuze, Oriane and Fernandes, Fernando Massa and Arib, Sofiane and Caputo, Laura and Fontes, Ana Pedro and Nguyen, Viet-Hung and Pazniak, Hanna and Nysten, Bernard and Charlier, Jean-Christophe and Hermans, Sophie and Hackens, Benoît},
        month = aug,
        year = {2025},
        pages = {186},
}

@article{Hu_2015,
        title = {Anisotropic electronic conduction in stacked two-dimensional titanium carbide},
        volume = {5},
        issn = {2045-2322},
        url = {https://www.nature.com/articles/srep16329},
        doi = {10.1038/srep16329},
        language = {en},
        number = {1},
        urldate = {2026-03-19},
        journal = {Scientific Reports},
        author = {Hu, Tao and Zhang, Hui and Wang, Jiemin and Li, Zhaojin and Hu, Minmin and Tan, Jun and Hou, Pengxiang and Li, Feng and Wang, Xiaohui},
        month = nov,
        year = {2015},
        pages = {16329}
}

@article{Celerier_2019,
	title = {Hydration of {Ti}$_{\textrm{3}}$ {C}$_{\textrm{2}}$ {T} \textit{$_{\textrm{x}}$ } {MXene}: {An} {Interstratification} {Process} with {Major} {Implications} on {Physical} {Properties}},
	volume = {31},
	issn = {0897-4756, 1520-5002},
	shorttitle = {Hydration of {Ti}$_{\textrm{3}}$ {C}$_{\textrm{2}}$ {T} \textit{$_{\textrm{x}}$ } {MXene}},
	url = {https://pubs.acs.org/doi/10.1021/acs.chemmater.8b03976},
	doi = {10.1021/acs.chemmater.8b03976},
	language = {en},
	number = {2},
	urldate = {2026-07-06},
	journal = {Chemistry of Materials},
	author = {Célérier, Stéphane and Hurand, Simon and Garnero, Cyril and Morisset, Sophie and Benchakar, Mohamed and Habrioux, Aurélien and Chartier, Patrick and Mauchamp, Vincent and Findling, Nathaniel and Lanson, Bruno and Ferrage, Eric},
	month = jan,
	year = {2019},
	pages = {454--461},
}

@article{Zaman_2021,
	title = {In situ investigation of water on {MXene} interfaces},
	volume = {118},
	issn = {0027-8424, 1091-6490},
	url = {https://pnas.org/doi/full/10.1073/pnas.2108325118},
	doi = {10.1073/pnas.2108325118},
	language = {en},
	number = {49},
	urldate = {2026-07-06},
	journal = {Proceedings of the National Academy of Sciences},
	author = {Zaman, Wahid and Matsumoto, Ray A. and Thompson, Matthew W. and Liu, Yu-Hsuan and Bootwala, Yousuf and Dixit, Marm B. and Nemsak, Slavomir and Crumlin, Ethan and Hatzell, Marta C. and Cummings, Peter T. and Hatzell, Kelsey B.},
	month = dec,
	year = {2021},
	pages = {e2108325118},
}

@article{Heinz_2013,
        title = {Thermodynamically {Consistent} {Force} {Fields} for the {Assembly} of {Inorganic}, {Organic}, and {Biological} {Nanostructures}: {The} {INTERFACE} {Force} {Field}},
        volume = {29},
        issn = {0743-7463},
        shorttitle = {Thermodynamically {Consistent} {Force} {Fields} for the {Assembly} of {Inorganic}, {Organic}, and {Biological} {Nanostructures}},
        url = {https://doi.org/10.1021/la3038846},
        doi = {10.1021/la3038846},
        number = {6},
        urldate = {2025-12-29},
        journal = {Langmuir},
        publisher = {American Chemical Society},
        author = {Heinz, Hendrik and Lin, Tzu-Jen and Kishore Mishra, Ratan and Emami, Fateme S.},
        month = feb,
        year = {2013},
        pages = {1754--1765},
}

@misc{Armstrong_2025,
        title = {Validated {Reactive} {Force} {Field} {Quantifies} {MXene} {Interfacial} {Properties}, {Mechanics}, {and Thermal} {Transport}},
        copyright = {https://creativecommons.org/licenses/by-nc/4.0/},
        url = {https://chemrxiv.org/doi/full/10.26434/chemrxiv-2025-212s7},
        doi = {10.26434/chemrxiv-2025-212s7},
        language = {en},
        urldate = {2026-02-17},
        author = {Armstrong, Isaac and Slocik, Joseph and Beck, Leo and Nepal, Dhriti and Zhu, Cheng and Varshney, Vikas and Heinz, Hendrik},
        month = dec,
        year = {2025},
}

@article{Mashtalir_2013,
        title = {Intercalation and delamination of layered carbides and carbonitrides},
        volume = {4},
        issn = {2041-1723},
        url = {https://www.nature.com/articles/ncomms2664},
        doi = {10.1038/ncomms2664},
        language = {en},
        number = {1},
        urldate = {2026-05-28},
        journal = {Nature Communications},
        author = {Mashtalir, Olha and Naguib, Michael and Mochalin, Vadym N. and Dall’Agnese, Yohan and Heon, Min and Barsoum, Michel W. and Gogotsi, Yury},
        month = apr,
        year = {2013},
        pages = {1716},
}

@article{Ferrara_2021,
        title = {The {Missing} {Piece}: {The} {Structure} of the {Ti}$_{\textrm{3}}$ {C}$_{\textrm{2}}$ {T}$_{\textrm{ \textit{x} }}$ {MXene} and {Its} {Behavior} as {Negative} {Electrode} in {Sodium} {Ion} {Batteries}},
        volume = {21},
        copyright = {https://creativecommons.org/licenses/by/4.0/},
        issn = {1530-6984, 1530-6992},
        shorttitle = {The {Missing} {Piece}},
        url = {https://pubs.acs.org/doi/10.1021/acs.nanolett.1c02809},
        doi = {10.1021/acs.nanolett.1c02809},
        language = {en},
        number = {19},
        urldate = {2026-05-28},
        journal = {Nano Letters},
        author = {Ferrara, Chiara and Gentile, Antonio and Marchionna, Stefano and Quinzeni, Irene and Fracchia, Martina and Ghigna, Paolo and Pollastri, Simone and Ritter, Clemens and Vanacore, Giovanni Maria and Ruffo, Riccardo},
        month = oct,
        year = {2021},
        pages = {8290--8297},
}

@article{Dong_2017,
        title = {Ti$_{\textrm{3}}$ {C}$_{\textrm{2}}$ {MXene}-{Derived} {Sodium}/{Potassium} {Titanate} {Nanoribbons} for {High}-{Performance} {Sodium}/{Potassium} {Ion} {Batteries} with {Enhanced} {Capacities}},
        volume = {11},
        issn = {1936-0851, 1936-086X},
        url = {https://pubs.acs.org/doi/10.1021/acsnano.7b01165},
        doi = {10.1021/acsnano.7b01165},
        language = {en},
        number = {5},
        urldate = {2026-05-28},
        journal = {ACS Nano},
        author = {Dong, Yanfeng and Wu, Zhong-Shuai and Zheng, Shuanghao and Wang, Xiaohui and Qin, Jieqiong and Wang, Sen and Shi, Xiaoyu and Bao, Xinhe},
        month = may,
        year = {2017},
        pages = {4792--4800},
}

@article{Wang_2015,
        title = {Atomic-{Scale} {Recognition} of {Surface} {Structure} and {Intercalation} {Mechanism} of {Ti}$_{\textrm{3}}$ {C}$_{\textrm{2}}$ {X}},
        volume = {137},
        issn = {0002-7863, 1520-5126},
        url = {https://pubs.acs.org/doi/10.1021/ja512820k},
        doi = {10.1021/ja512820k},
        language = {en},
        number = {7},
        urldate = {2026-05-28},
        journal = {Journal of the American Chemical Society},
        author = {Wang, Xuefeng and Shen, Xi and Gao, Yurui and Wang, Zhaoxiang and Yu, Richeng and Chen, Liquan},
        month = feb,
        year = {2015},
        pages = {2715--2721},
}

@article{Benchakar_2020,
        title = {One {MAX} phase, different {MXenes}: {A} guideline to understand the crucial role of etching conditions on {Ti3C2Tx} surface chemistry},
        volume = {530},
        issn = {01694332},
        shorttitle = {One {MAX} phase, different {MXenes}},
        url = {https://linkinghub.elsevier.com/retrieve/pii/S0169433220319668},
        doi = {10.1016/j.apsusc.2020.147209},
        language = {en},
        urldate = {2026-05-28},
        journal = {Applied Surface Science},
        author = {Benchakar, Mohamed and Loupias, Lola and Garnero, Cyril and Bilyk, Thomas and Morais, Cláudia and Canaff, Christine and Guignard, Nadia and Morisset, Sophie and Pazniak, Hanna and Hurand, Simon and Chartier, Patrick and Pacaud, Jérôme and Mauchamp, Vincent and Barsoum, Michel W. and Habrioux, Aurélien and Célérier, Stéphane},
        month = nov,
        year = {2020},
        pages = {147209},
}
%


\end{document}